\documentclass[conference]{IEEEtran}
\usepackage{cite}
\usepackage{amsmath,amssymb,amsfonts}
\usepackage{algorithmic}
\usepackage{graphicx}
\usepackage{textcomp}
\usepackage{xcolor}
\usepackage{multirow,booktabs,makecell}
\usepackage{verbatim}
\usepackage[vlined,boxed,ruled]{algorithm2e}
\def\BibTeX{{\rm B\kern-.05em{\sc i\kern-.025em b}\kern-.08em
    T\kern-.1667em\lower.7ex\hbox{E}\kern-.125emX}}
\begin{document}

\title{LVMapper: A Large-variance Clone Detector Using Sequencing Alignment Approach\\
}

\author{\IEEEauthorblockN{Ming Wu\IEEEauthorrefmark{1}\IEEEauthorrefmark{2}, Pengcheng Wang\IEEEauthorrefmark{1}\IEEEauthorrefmark{2}, Kangqi Yin\IEEEauthorrefmark{1}\IEEEauthorrefmark{2}, Haoyu Cheng\IEEEauthorrefmark{1}\IEEEauthorrefmark{2}, Yun Xu\IEEEauthorrefmark{1}\IEEEauthorrefmark{2} and Chanchal K.Roy\IEEEauthorrefmark{3}}
\IEEEauthorblockA{\IEEEauthorrefmark{1}School of Computer Science, University of Science and Technology of China, Hefei, China}
\IEEEauthorblockA{\IEEEauthorrefmark{2}Key Laboratory on High Performance Computing, Anhui Province, Hefei, China\\
Email: \{wuming, wpc520, yinkq, chhy\}@mail.ustc.edu.cn, xuyun@ustc.edu.cn\\
Corresponding author: Yun~Xu (xuyun@ustc.edu.cn)}
\IEEEauthorblockA{\IEEEauthorrefmark{3}Department of Computer Science, University of Saskatchewan, Saskatoon, Canada \\
Email: chanchal.roy@usask.ca}
}

\maketitle

\begin{abstract}
To detect large-variance code clones (i.e. clones with relatively more differences) in large-scale code repositories is difficult because most current tools can only detect almost identical or very similar clones. It will make promotion and changes to some software applications such as bug detection, code completion, software analysis, etc. Recently, CCAligner made an attempt to detect clones with relatively concentrated modifications called large-gap clones. Our contribution is to develop a novel and effective detection approach of large-variance clones to more general cases for not only the concentrated code modifications but also the scattered code modifications. A detector named LVMapper is proposed, borrowing and changing the approach of sequencing alignment in bioinformatics which can find two similar sequences with more differences. The ability of LVMapper was tested on both self-synthetic datasets and real cases, and the results show substantial improvement in detecting large-variance clones compared with other state-of-the-art tools including CCAligner. Furthermore, our new tool also presents good recall and precision for general Type-1, Type-2 and Type-3 clones on the widely used benchmarking dataset, BigCloneBench.
\end{abstract}


\section{Introduction}

Clone code is generated by copying, pasting and modifying code fragments for reuse, which are common operations in software development \cite{roy2007survey, koschke2007survey}. In the past, code clones with relatively more modifications (called large-variance code clones) were difficult to be found by existing tools \cite{wang2018ccaligner} because most of them were suitable for finding the identical or very similar clones. However, in our experimental observation, large-variance code cloning is ubiquitous. Compared with the clones provided by traditional tools, the large-variance code clones have a broader range. Accordingly, they have an important impact on and make changes to some software applications such as bug detection, code completion, software analysis and so on. Recently, a meaningful attempt has been made with CCAligner \cite{wang2018ccaligner}. It is a large-gap clone detection tool which can detect the code clones with relatively concentrated modifications. Our study focuses on a more general case: to detect large-variance code clones that include not only the clones with concentrated code modifications but also those with scattered code modifications.

\begin{figure*}[htbp]
\centerline{\includegraphics[width=0.95\textwidth]{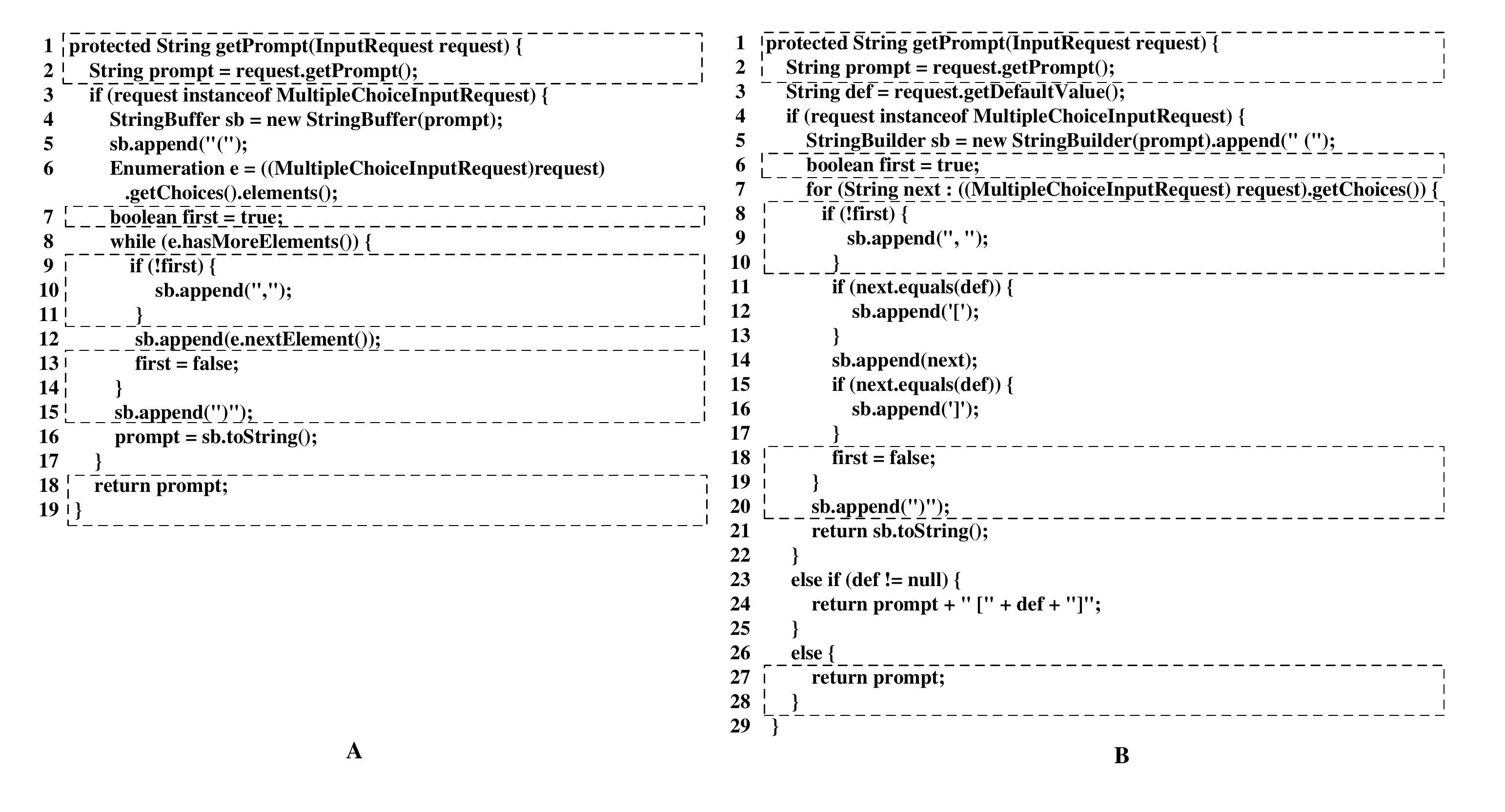}}
\caption{Example of a large-variance clone.}
\label{fig1}
\end{figure*}

Software development usually involves two modes: \emph{homologous modification} and \emph{heterologous development}. In \emph{homologous modification} there is always an original version of code and modifications on the original version generate new clone code. Hence, the similarity in lexical is the inherent characteristic of \emph{homologous modification}. By contrast, in \emph{heterologous development}, different programmers develop similar functionalities and these introduce semantic clones, where the lexical similarity is not inherent. Furthermore, large-variance clones of \emph{homologous modification} are generated by two cases. One of the common cases is that the old version of software are modified and expanded iteratively. Another case is the reuse and modification of open source software. Both cases are very common in software development, therefore it is important to detect the clones with \emph{homologous modification}. Fig.~\ref{fig1} shows an example of clones between two different versions of project \emph{Ant 1.6.5} and \emph{1.10.5}. In code B, the code segments of lines 1--2, 6, 8--10, 18--20, 27--28 are the same as those of the lines 1--2, 7, 9--11, 13--15, 18--19 in code A, respectively. Lines 4--5, 7, 14 and 21 in code B are modified from lines 3--5, 8, 12 and 16 in code A. Other lines in code B are new extension part of code A. These modifications in clone codes are scattered and occupy a certain portion of the code.

Existing tools have made attempts but still have more or less limitations in finding Type-3, especially large-variance code clones. For one of the best tools with good performance on Type-3 clone, SourcererCC \cite{sajnani2016sourcerercc} has to decrease the threshold of similarity to find large-variance clones at the cost of accuracy loss. Another popular detector CCFinderX is good at identifying Type-1 clones and Type-2 clones, but cannot directly support Type-3 clone detection. iClones identifies the Type-3 clones by merging the nearby small Type1-2 clone fragments, but the recall of Type-3 is low due to the simple strategy. NiCad uses a Longest Common Sub-sequence (LCS) algorithm and can tolerate discontinuous subsequences. However, it does not scale and its precision suffers with decreasing thresholds. CCAligner is a good recent attempt in detecting clones with relatively concentrated modifications. It can detect large-gap but misses scenarios where modifications are scattered. Besides, some semantic methods have certain ability to detect variance clones because there is an overlap between semantic clones and syntactical clones. Deckard \cite{jiang2007deckard} builds the characteristic vectors from abstract syntax tree (AST) to detect clones, but suffers from low precision and recall rate. Deep learning methods such as Oreo \cite{saini2018oreo} encode software metrics into semantic vectors and achieve good results, but they mainly focus on semantic clones.

For these considerations, we present a tool aimed at detecting large-variance code clones called LVMapper. Our proposed code clone detector that can find clones with more general variance is based on locate-filter-verify method. Its key idea mainly comes from third-generation sequencing alignment method \cite{altschul1990basic, liu2015rhat, li2010survey}. In bioinformatics, the third-generation sequencing alignment based on seed-and-extend strategy performs well with sequence difference up to 30\%. LVMapper uses small windows of continuous lines (called \emph{seeds}) with lower costs to locate and filter the candidate pairs of clone codes. In order to verify whether these candidate pairs of codes are cloned, another feature that code clones always have certain proportion of order-preserving code lines is considered. Based on this property, a heuristic algorithm which is more efficient than the Longest Common Subsequence (LCS) algorithm is proposed. Besides, a dynamic threshold that changed with the code size is used for the verification of code clones. It makes LVMapper identify clones with more modifications while guaranteeing certain precision.

To evaluate the large-variance clone detection performance of LVMapper, we carried out experiments on self-synthetic and real datasets. In order to test the capability of finding large-variance clone, we compared our tool's performance with NiCad, SourcererCC and CCAligner on 4 Java and 4 C projects. For the self-synthetic dataset experiment, we generated clones by inserting scattered different lines to source code, and then evaluated the performance of LVMapper and other tools in detecting clones with various modifications. We also used the BigCloneBench \cite{svajlenko2014towards, svajlenko2015evaluating} to compare and measure the different type clones recall of LVMapper with CCFinderX \cite{kamiya2008official}, iClones \cite{gode2009incremental}, Deckard \cite{jiang2007deckard}, NiCad \cite{roy2008nicad}, SourcererCC \cite{sajnani2016sourcerercc} and CCAligner \cite{wang2018ccaligner}. The experiments show that LVMapper performed the best in detecting large-variance clones and had good recall and precision for general Type-1 to Type-3 clones.

The main contributions of our work are as follows:

(1)	Goal contribution: CCAligner has advantages in detecting large-gap clones while our work extends the detection approach of large-variance clones to more general cases. It identifies not only the clones with concentrated code modifications but also the clones with the scattered code modifications. We also give a concrete definition of the large-variance clones. 

(2)	Method contribution: Inspired by the idea of the seed-and-extend method in bioinformatics, we develop a novel tool with locate-filter-verify procedure and it is suited to detect clone with large variance. We propose a dynamic threshold to promote the accuracy and recall, and a rapid method to verify, avoiding the time-consuming dynamic programming.

(3)	Result contribution: We compared LVMapper with other state-of-the-art detectors on real cases of software projects, self-synthetic programs and the state-of-the-art benchmarks. The results show that LVMapper is much better than the state-of-the-art tools in large-variance clone detection. In addition, our new tool has good recall and precision for general Type-1, Type-2 and Type-3 clones.


The rest sections of the paper are organized as follows. Some terminologies and definitions of code clone are introduced in Section \ref{Definition}. Section \ref{Method} provides the details of our detection tool. Section \ref{Evaluation} presents the results of the experiments to evaluate the detection ability of our approach. In Section \ref{Related Work} the related work of clone detection is discussed and Section \ref{Limitation} describes the limitation. Finally, Section \ref{Conclusion} concludes the paper and briefly introduces future work.

\section{Terminologies \& Definitions}\label{Definition}

\emph{Code block} is a statement sequence within braces and usually represents a single function. \emph{Clone pair} is a pair of similar code portions. The \emph{minimum clone size} is the minimum number of lines or tokens that either of a clone pair should have. The standard minimum clone size is 6 lines or 50 tokens which we also follow in this paper. 
Four primary clone types are agreed by researchers and the former work \cite{bellon2007comparison, roy2007survey}:

\emph{Type-1 (textual similarity)} and \emph{Type-2 (lexical similarity)} clones are syntactically identical code fragments except for variances in white space, layout, comments and variances in identifier names, literal values, white space, layout and comments, respectively. \emph{Type-3 (syntactic similarity)} clones are code fragments which are similar but have statements added, modified and/or removed with respect to each other. \emph{Type-4 (semantic similarity)} clones are code fragments that implement the same functionality but are different in syntax.

Type-3 and Type-4 clones are difficult to partition because there is no clear boundary between syntactically similar Type-3 clone and dissimilar Type-4 clone. Hence, BigCloneBench \cite{svajlenko2014towards} further divided Type-3 and Type-4 into four types according to the syntactical similarity range: Very Strong Type-3 similarity in range [0.9, 1.0), Strongly Type-3, [0.7, 0.9), Moderately Type-3, [0.5, 0.7), and Weakly Type-3\&4, [0.0, 0.5). 

Before giving the definition of large-variance clones, we first define a new similarity of code pairs and then obtain the difference degree of code pairs. In the past, Jaccard similarity \cite{jaccard1912distribution} is usually used to measure the similarity of code pairs, which is not suitable for code pairs with large difference in size. For example, given code block \emph{A} and \emph{B}, assume the size of \emph{B} is double the size of \emph{A} and all lines of \emph{A} appears in \emph{B}. According to the Jaccard similarity, the similarity of code blocks \emph{A} and \emph{B} are only 0.5, which is not practical. In fact, all of \emph{A} are cloned by \emph{B}.

\emph{Definition 1  Similarity and Difference of code pairs}: Given two code blocks \emph{A} and \emph{B} consisting of \emph{$l(A)$} and \emph{$l(B)$} pretty-printed lines, respectively. Let \emph{$l(A \cap B)$} be the number of shared lines in \emph{A} and \emph{B}. The harmonic similarity \emph{sim}($A,B$) of \emph{A} and \emph{B}, the single-side similarity \emph{sim}($A|B$) that \emph{A} relative to \emph{B}, and the single-side similarity \emph{sim}($B|A$) that \emph{B} relative to \emph{A} are:
\begin{equation}
sim(A,B) =  \frac{1}{2} \cdot \left( \frac{l(A \cap B)}{l(A)}+\frac{l(A \cap B)}{l(B)} \right) \label{eq1}
\end{equation}

\begin{equation}
sim(A|B) = \frac{l(A \cap B)}{l(A)}, sim(B|A) = \frac{l(A \cap B)}{l(B)}
\label{eq2}
\end{equation}

The difference \emph{\text{diff(A,B)}} of \emph{A} and \emph{B},and the difference \emph{diff}($A|B$) that \emph{A} relative to \emph{B}, and the difference \emph{diff}($B \mid A$) that \emph{B} relative to \emph{A} are defined according to the similarity, respectively:

\begin{equation}
\emph{diff}(A,B) =  1-sim(A,B) \label{eq3}
\end{equation}
\begin{equation}
\emph{diff}(A | B) =  1-sim(A | B), \emph{diff}(B | A) = 1-sim(B | A)
\label{eq4}
\end{equation}

Taken the previous example of code \emph{A} and code \emph{B} which is double size of \emph{A}, the single-side similarity $\emph{sim}(A|B)= 1$, which means that \emph{A} is all in \emph{B} and it matches the actual situation. The harmonic similarity $\emph{sim(A,B)} = 1/2 \cdot (\frac{m}{m} + \frac{m}{2m}) = 0.75$, where \emph{m} is  the number of lines in code \emph{A}, is greater than the Jaccard similarity. On these bases, we give the definition of large-variance clone as follows.

\emph{Definition 2 Large-variance clone}: If code blocks \emph{A} and \emph{B} are clone and \emph{\text{diff(A,B)}} $> \Delta$ (or $\emph{diff}(A|B) > \Delta$, or $ \emph{diff}(B|A) > \Delta$), then \emph{A} and \emph{B} are $\Delta$-difference clone ($0 \le \Delta \le 1$). Particularly, for \emph{\text{diff(A,B)}} $\Delta$ is set as 0.15, i.e. \emph{\text{diff(A,B)}} $> 0.15$, then \emph{A} and \emph{B} are called large-variance clone (abbreviated as \emph{LV clones}).

The difference degree of large-variance clones is set higher than 0.15 mainly based on two considerations. First, it's difficult for most code clone detection tools or methods to find such clones, even the better tools that are able to detect Type-3 clones cannot find them well. Another  consideration is the compatibility with the large-gap clones described in \cite{wang2018ccaligner}, in which the volume difference between blocks \emph{A} and \emph{B} is used to specify variance clones. Assuming that \emph{A} is the block with smaller size and \emph{B} is the block with larger size, their method is to detect the large-gap clones for the set of blocks $\lambda$-difference volume ($\lambda = l(A) / l(B) \le$ 0.7). According to our definitions, the \emph{\text{sim(A,B)}} is at most 0.85, then the \emph{\text{diff(A,B)}} is at least 0.15. Hence, the code clones described in \cite{wang2018ccaligner} are included in our large-variance clones but ours have a broader scope.

\section{Method}\label{Method}

\begin{figure}[htbp]
\centerline{\includegraphics[width=0.54\textwidth]{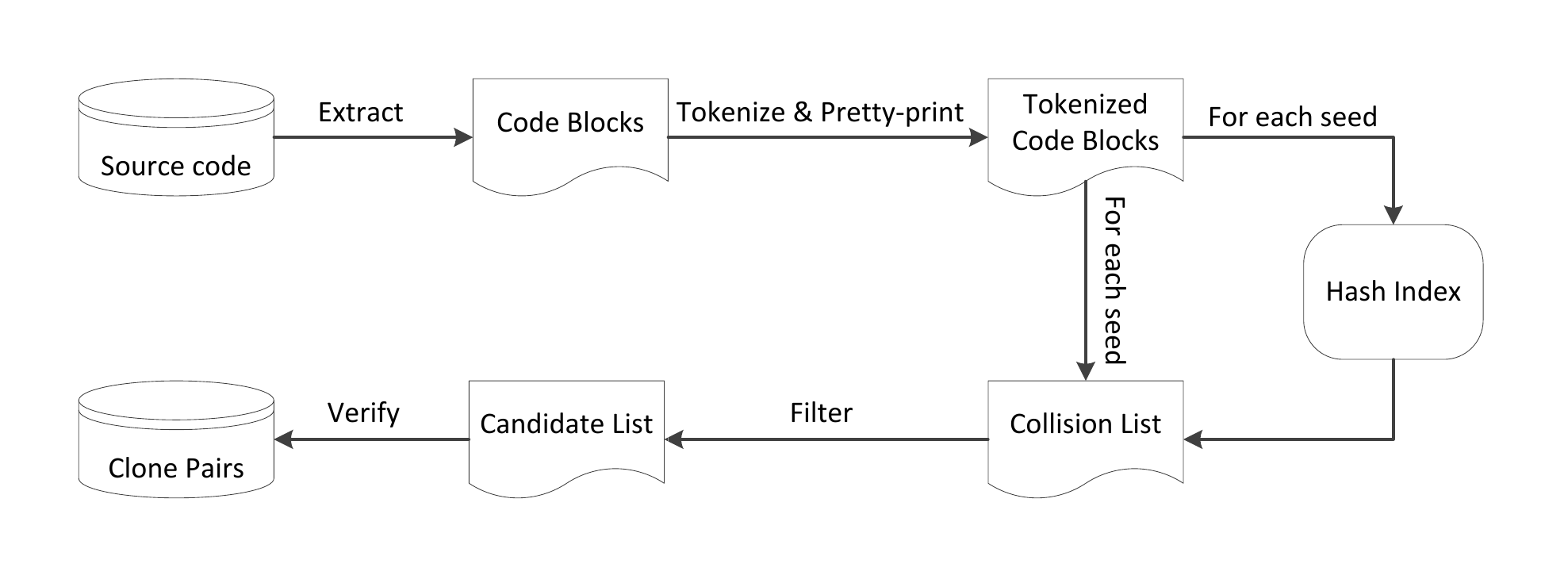}}
\caption{General procedure of LVMapper.}
\label{fig2}
\end{figure}

Inspired by the seed-and-extend approach which is typically used in sequencing alignment from bioinformatics \cite{altschul1990basic, liu2015rhat, cheng2015bitmapper}, we proposed a locate-filter-verify procedure for clone detection. In bioinformatics, the seeding step uses the subsequences of the query to quickly locate exact match in reference and the extending step extends and refines the candidate positions by a dynamic programming alignment. Our method includes three phases: locate-filter-verify. The first two phases are designed to seek out the candidate clone pairs with low cost and high recall. In the last phase we design a heuristic algorithm to further eliminate the false clone pairs and improve the accuracy. The uses of dynamic threshold, seeds index and avoiding time-consuming dynamic programming are the keys and innovations of our method. Fig.~\ref{fig2} shows the general procedure. The rest of this section will provide detail descriptions of each phase.

\subsection{Lexical Analysis}

Lexical analysis for code includes extracting code blocks from source code and tokenizing the code. TXL \cite{cordy2006txl}, which is commonly used in previous tools \cite{wang2018ccaligner, sajnani2016sourcerercc}, is adapted to extract code blocks from source code files. After obtaining the code blocks, the tokenizing step that mostly based on Flex \cite{paxson1995flex} begins. Identifiers including variables and function names are replaced by the same token `id' to tolerate Type-2 changes. The extracted code blocks are pretty-printed. The tokens of each line are concatenated into a single token sequence except the white spaces.

\subsection{Seeds Indexing}

It is necessary to establish a seed index to speed up the computation for the locating and filtering phases, where the \emph{seeds} are all of the \emph{k}-line sliding windows (i.e., code fragments of continuous \emph{k} lines) for a code block. The seeds are basic units for matching instead of tokens. For example, given a code block with 10 lines and sliding windows size \emph{k} = 3, the number of all seeds is obviously 8. In LVMapper, these seeds are converted to hash value and the seed is also regarded as the hash value.

Here are the detailed steps of indexing. LVMapper scans all code blocks, collects all seeds and indexes them in a hash table. The key of the element in hash table is seed's hash and the value is a set of corresponding positions (e.g., block number plus line number). Like CCAligner, we use MurmurHash hash function \cite{appleby16} in order to guarantee the efficiency with the low collision rate. For the value in the hash table, we packed the block number and line number into a 64-bit integer. 

\subsection{Locating via the Shared Seed}\label{Locating}

The locating phase is a preliminary selection for possible code clone pairs. In this phase, the goal is to collect as many candidates as possible without losing real clone pairs. Because the standard minimum clone size is 6 lines \cite{svajlenko2014towards, bellon2007comparison}, CCAligner chooses 6-line or longer windows to match the possible clone pairs. The experiments in CCAligner \cite{wang2018ccaligner} shows that 6 lines with 1 mismatch windows balanced recall with precision. However, the cost of this approach is still considerably high. We use a lower cost and more efficient way to achieve this. Here, the 3-line sliding windows are chosen as seeds to collect the possible clone pairs that share seed(s).

For any code segments of 6-line with 1 mismatch that can be found by CCAligner, in LVMapper, they can also be identified by two non-overlapping 3-line exact windows. The reason is that the 6 lines window in CCAligner can be covered by two non-overlapping 3-line windows. According to pigeonhole principle, one line modification affects at most one 3-line window and the other one remains unchanged. As a result, the identification ability of our method is better than that of CCAligner.  

Algorithm~\ref{Alg:1} lists the steps of clone detection process, in which lines 4--11 belong to the locating phase. To retrieve these seeds, we create a hash table for efficiency. Once the index has been built, the candidates of each code clock can be obtained by utilizing this index. Let the current inquiring code block be \emph{A}. Every sliding overlapping 3-line window in \emph{A} is the seed, whose hash value is used to find positions in hash table with the same key (\emph{line 9}). If the block id of the position, denoted as \emph{B}, is greater than \emph{A}, the position will be added to \emph{CollisionList (A)} the collision list of block \emph{A} (\emph{line 11}). This practice eliminates the duplication of detecting clone for the same two blocks with reverse order. When \emph{B} is the current inquiring block, block \emph{A} is not considered anymore because the pair of \emph{A} and \emph{B} has been considered before. After the last seed of current block is queried, the positions in the \emph{CollisionList (A)} are sorted according to block id and then according to line id (\emph{line 12}). Every block \emph{B} that has the collided seed with \emph{A} will be further filtered and verified.

\begin{algorithm}[!htbp]
\scriptsize
\SetCommentSty{small}
\LinesNumbered
\caption{Clone Detection}
\label{Alg:1} 
\KwIn{$A$ is a list of tokenized code blocks \{$a_1$, $a_2$,...$a_n$\}, Hash Table $H$ of $A$, window size k, threshold $\theta$ for filtering phase, threshold $\delta$ for verifying phase}
\KwOut{All clone pairs $CP$}
$H \gets \varnothing$\;
$CP \gets \varnothing$\;
$len = \text{number of lines in $a_i$}$\;
\For{each $a_i$ in $A$}  
{  
	\tcc{Locating phase}
	\For{$j=1;j\le len-k+1;j++$}
     {  
     		$l_j = a_i.linej$\;
		$win_j = \text{CONCAT}(l_j,l_{j+1},...l_{j+k-1})$\;
		$key = \text{HASH}(win_i)$\;
		$pos = \text{FIND}(H,key)$\;
		\If{$pos > a_i$}
		{
			$CollisionList_i = CollisionList_i \cup pos$\;
		}
		
     } 
	SORT($CollisionList_i$)\;
	\For{each block $B$ in $CollisionList_i$}
	{
		\tcc{Filtering phase}
		$s =  \text{number of $B$ in $CollisionList_i$}$\;
		$L =$ number of lines in $B$\;
		$SR = s / (L-k+1)$\;
		\If{$SR \ge \theta$}
		{
			\tcc{Verifying phase}
			$block1 = $block with smaller size between $a_i$ and $B$\;
			$block2 = $block with larger size between $a_i$ and $B$\;
			$line1 = 1$\;
			$comm\_lines = 0$\;
			$lastline = 0$\;
			\While{$line1 \ne block1.end$}
			{
				$k1 =$ HASH($block1.line1$)\;
				\tcc{FIND\_IN\_BLOCK find the first line after $lastline$ in $block2$ and has the same hash value of $k1$}
				$line2 =$ FIND\_IN\_BLOCK($k1,block2,lastline$)\;	
				\If{$line2 \ne NULL$}
				{
					$seglen = 0$\;
					$ m = 1$\;
					\While{$block1.line1+m = block2.line2+m$}
					{
						$seglen++$\;
						$m++$\;
					}
					\If{$seglen \ge 2$}
					{
						$comm\_lines += seglen$\;
					}
				}
				$line1++$\;				
			}
			$min\_lines =$  minimum lines of $a_i$ and $B$\;
			$OS = comm\_lines / min\_lines$\;
			\If{$OS \ge \delta$}
			{
				$CP  = CP \cup (a_1,B)$\;
			}

		}

	}

}
return $CP$\;
\end{algorithm}

\subsection{Filtering via the Common Seeds Number}\label{Filtering}

Through the locating phase, two code blocks that share common seed(s) may be clone pair. Then we take into account the clone possibility of these \emph{Collision} code blocks. This phase is called the filtering phase and it picks out candidate clone pairs. It gives us two benefits. First, the number of candidate code pairs can be reduced significantly. Therefore the processing time and the false positive rate can also be reduced. Besides, the probability of the candidate pairs being the true clones increase significantly. Our idea of the filtering phase is mainly based on considering the number of shared seeds for two code blocks to measure the possibility of being clone.

The selection of candidate clone pairs depends on the similarity between the two code blocks, and the similarty is calculated by the number of seeds they share. In our method, we adopt the Equation \eqref{eq2} to compute the similarity of a code pair, because the definition of single-side similarity has better ability to discover large variance code clones. As we get the number of shared seeds between two code blocks, the former of Equation \eqref{eq2} can be equivalently changed to:
\begin{equation}
SR_A(B) =  \frac{s}{t} = \frac{s}{L - k + 1} \label{eq5}
\end{equation}
where \emph{s} is the number of shared seeds, \emph{t} is total seeds number of \emph{B} and \emph{L} is the length in line of \emph{B}. For any pair of the collision code blocks, the higher the \emph{$SR_A(B)$} value is, the more likely they are to be a clone pair.

The method of threshold setting for \emph{$SR_A(B)$} is the key point of our method, and it is also significantly different from other methods. In order to enhance the ability of detecting large variance code clones, here we use a dynamic threshold to filter the code pairs. The function of the threshold is designed as a multi-segment function, corresponding to the length of code blocks A and B. As the length of code blocks A and B increases, this threshold can be reduced. Actually, the reason can be explained by an understandable analogy. For example, given real-life conversations, how to judge whether two different conversations belong to the same topic? We have such a consensus that long conversations with lower rate of common sentences could discuss the same subject while short conversations should have higher rate to judge as discussing the same subject. In the experiments of next section, the filtering threshold is set to about 0.1 for both code blocks with lengths more than 10. Note that SourcererCC uses the fixed ratio of shared tokens to verify clone pairs and it sets the ratio threshold as 0.7 to guarantee the precision.

Lines 13--16 in Algorithm~\ref{Alg:1} belongs to filtering phase. In practice, once we get the candidate blocks list \emph{CollisionList (A)} of current inquiring block \emph{A}, for every candidate block \emph{B} in the list, LVMapper counts the number of \emph{B} in \emph{CollisionList (A)} (\emph{line 14}). As we mentioned above, we treat each overlapping \emph{k}-line windows as seed to vote for potential clone blocks, then every position added in the collision list is the block that have the same seed with \emph{A}. In this case, the number \emph{B} in \emph{CollisionList (A)} is the number of votes that \emph{B} gets from \emph{A}. If \emph{B} gets more votes, then it is more likely to be clone code of \emph{A}. The idea is similar to the idea of seed-and-extend in the sequencing alignment \cite{liu2015rhat}. The threshold for \emph{$SR_A(B)$} is $\theta$ (\emph{line 17}).

\subsection{Verifying via the Ordered Common Lines}

Unlike the first two phases, the last phase (called the verifying phase) further measures the clone possibility of the candidate clone pairs output by the filtering phase from another perspective: if two blocks are large-variance clone, an important feature of them is that the common lines of code in them have order preserving property. Actually, previous tool NiCad \cite{roy2008nicad} used similar idea. It is based on a Longest Common Sub-sequence (LCS) algorithm. Not as complex as NiCad, we design a heuristic simple algorithm for this order preserving property. The idea of the heuristic algorithm is to count the order preserving number of two adjacent code lines in one code block.

Similarly, the similarity of the candidate pair is measured by another characteristic quantity: the rate of ordered common lines. This characteristic quantity \emph{OS (A, B)} is defined as: 
\begin{equation}
OS (A, B) =  \frac{comm\_lines}{min\_lines(A, B)} \label{eq6}
\end{equation}
where \emph{common\_lines} is the ordered common lines of \emph{A} and \emph{B}, and \emph{min\_lines(A, B)} is the minimum size in line of \emph{A} and \emph{B}. Like the threshold mentioned in the filtering phase, we also use a dynamic threshold for verifying candidate clone pairs. The detailed setting will also be discussed in Section \ref{Evaluation}.

We implement the heuristic algorithm as follows and lines 17--38 in Algorithm~\ref{Alg:1} show the steps. For every candidate pair of block \emph{A} and \emph{B} survived from the locating and filtering phase, assume block \emph{A} is smaller than \emph{B}. We scan from the first line in \emph{A} to find contiguous lines which also appear in \emph{B} (\emph{line 25}). LVMapper keeps a variable \emph{comm\_lines} to record the sum of matching lines. If there are at least 2 contiguous shared lines, then the length is added to \emph{comm\_lines} (\emph{line 33}). The contiguous lines of A and B are in order and the segments of the contiguous lines are also in order. The threshold $\delta$ of \emph{OS (A, B)} to verify candidate pairs (\emph{line 37}) is set according to the minimum size of \emph{A} and \emph{B}.

\section{Evaluation}\label{Evaluation}
The performance of LVMapper for detecting large-variance clones and general Type-1, Type-2, Type-3 were thoroughly evaluated in real and self-synthetic dataset. We first introduce the parameter setting of seed length and dynamic threshold of different phases. Then the detailed information of different experiments is provided.

\subsection{Parameter Setting}

\subsubsection{Choice of Seed Length}

As the seeds play a major role in locating and filtering phases, the choice of seeds length is important and has an impact on the performance of LVMapper. If the seeds are too long, the recall of our method will be affected. In contrast, if the seeds are too short, the effectiveness of the locating and filtering will be eroded. In Section \ref{Method} we analyzed the choice of seed length theoretically. Here we also used experiments to evaluate the performance of detection for different seed lengths \emph{k} quantificationally. We used the BigCloneBench \cite{svajlenko2014towards, svajlenko2015evaluating} to evaluate the detection ability of LVMapper for different seed lengths, because it is not only a benchmark for general clones but also contains large-variance clones. Besides, we considered the memory use and execution time of different seed lengths for Linux kernel dataset.

BigCloneBench is a benchmark which contains different types of manually validated clones in the repository IJaDataset-2.0 \cite{ambient13} and it defines clone types by syntactic similarity as described in Section \ref{Definition}. The framework BigCloneEval \cite{svajlenko2016bigcloneeval} summarizes recall performance for different clone types of clone detectors automatically and it is widely used in previous work \cite{sajnani2016sourcerercc, saini2018oreo}.  We configured the BigCloneEval with minimum clone size 6 lines and 50 tokens which are consistent with the standard minimum clone size. The seed length of LVMapper was set as 2-line, 3-line and 4-line, with other parameters fixed. The recall was reported by BigCloneEval. And for each parameter, we measured the precision by randomly validating 400 reported clone pairs.

Table~\ref{tab1} shows the detailed results. Because the recall rate of Weakly Type-3\&4 is under 1\%, we provided the number of detected clones instead, denoted as \emph{\# of Weakly Type-3\&4}. As seen from Table~\ref{tab1}, the recall of Type-1, Type-2 and Very Strongly Type-3 were nearly 100\% for all seed length. When the seed length became longer, the recall of type-3 with lower similarity decreased. For seed length of 4-line, the recall of Strongly Type-3 and Moderately Type-3 was under 80\% and 20\%, respectively. And the number of Weakly Type-3\&4 fell to 22998. However, while the recall for seed length of 2-line was the highest, especially in Weakly Type-3\&4, the precision declined apparently. The recall and precision for seed length 3-line strike a balance. The number of Weakly Type-3\&4 was over 30000 and the precision kept at 88\%.

\begin{table}[htbp]
\renewcommand\arraystretch{1.4}
\caption{Recall per Clone Type and Precision Measured for BigCloneBench with Different Seed Length}
\begin{center}
\resizebox{0.8\columnwidth}{!}{%
\scriptsize
\begin{tabular}{cccccccc}
\toprule[0.6pt]
k  	 	&2   		& 3    		& 4  		\\
\midrule
Type-1 	& 100  	& 100      	& 100		\\

Type-2 	& 99   	& 99         	& 99 		\\

Very Strongly Type-3 & 98 & 98  	& 98  	\\

Strongly Type-3 & 85 & 81 			& 77 		\\

Moderately Type-3 & 21 & 20   		& 18  	\\

\# of Weakly Type-3\&4 & 35613 & 30250  & 22998  \\

\midrule

Precision & 85 		& 88       	&  89   \\ 

\bottomrule[0.6pt]
\end{tabular}
}%
\label{tab1}
\end{center}
\end{table}

Besides, we took into consideration the memory use and execution time of different seed lengths. The Linux kernel 4.18 was used as the target source code and it has 25782 files with 12964738 lines of code (LOC) measured by cloc \cite{cloc15}. As shown in Table~\ref{tab2}, the execution time of 2-line method was as much as 19 times compared to the execution time of 3-line method and the memory use increased about 100MB. The configuration of seed length 3-line had the least memory use and medium execution time. The execution time of 4-line method was the shortest, but it had more memory requirement. Note that the configuration of seed length 3-line has the least memory use. The reason is that for the method using 4-line, larger window space allows greater variation of seed, resulting in greater hash index space. For the method using 2-line, more position values (i.e., the block id and line id corresponding to the seeds) occupy the storage space. Taken together, the seed length of 3-line balanced not only the recall and precision, but also the space and time. Therefore, we selected 3-line as our default configuration.

\begin{table}[htbp]
\caption{Execution Time and Memory Space with Different Parameterization for Linux 4.18}
\begin{center}
\resizebox{0.8\columnwidth}{!}{%
\scriptsize
\begin{tabular}{cccc}
\toprule[0.6pt]

 k & 2 & 3 & 4 \\ 
\midrule
Time & 11h 28m 48s  & 36m 16s  &  13m 32s \\ 
Memory & 841 MB & 760 MB & 834 MB \\

\bottomrule[0.6pt]
\end{tabular}
}%
\label{tab2}
\end{center}
\end{table}

\subsubsection{Threshold for Filter and Verification}
In filtering and verifying phases, we use dynamic threshold for judgment of results. The threshold is defined according to the block size in order to be better adapted to code clone judgment. 

For $\theta$, the threshold of \emph{$SR_A(B)$} in the filtering phase, it is defined as:
\begin{equation}
\theta =
\begin{cases}
\mu*L + \nu & \text{if } 6 < L \le 10,\\
0.1 & \text{if } L > 10.
\end{cases} \label{eq7}
\end{equation}
where \emph{L} is the size in length of the collided block \emph{B}. We set $\theta$ = 0.5 when \emph{B} has 6 lines and we set $\theta$ = 0.1 when the size of \emph{B} is greater than 10 lines. When the code size is small, there are plenty of statements that have similar forms. So LVMapper filters the smaller candidate block with stricter standards. The use of the dynamic threshold utilizes the characteristic of source code.

Moreover, for the ratio of ordered common sequences in the verifying phase, assume the smaller block of the candidate pair is block \emph{A}. We adapt a more-refined piecewise function to define the threshold $\delta$, which is used in the verifying phase, according to \emph{A.size} \emph{l}:
\begin{equation}
\delta =
\begin{cases}
0.55  & \text{if } 6 < l \le 10,\\
g(l)  & \text{if } 10 < l \le 20,\\
0.3  & \text{if } l > 20.
\end{cases} \label{eq8}
\end{equation}
In Equation \eqref{eq8}, g(l) = -$\alpha$$\cdot$\emph{l}+$\beta$ relies on size of \emph{A}. In our implementation, we set $\alpha$ = 0.025, $\beta$ = 0.8 empirically. For the smaller blocks whose length is smaller than 10 lines, the ratio of minimum matching continuous lines is 0.55, because in our observation the precision will be slashed when $\delta$ $<$ 0.55. For medium size blocks with length from 10 to 30 lines, the ratio of ordered common sequences linearly decreases with the smaller blocks length. As big blocks are more likely to be modified in code clone, the lower limit of $\delta$ is 0.3 which allows large-variance for big code blocks and ensures certain accuracy.

\subsection{Large-variance Clone Detection} \label{LvDetect}
To test the large-variance clone detection ability, we first compared LVMapper with others in eight open source projects dataset. Then we tried to construct synthetic dataset by inserting different number of lines to further evaluate the detecting ability for different variance proportions. 

\subsubsection{Empirical Test} \label{Empirical}

Here we evaluated the large-variance clone detection ability and studied the existence and pervasiveness of large-variance clones. Considering that CCAligner is a good large-gap clone detection method in a very recent study \cite{wang2018ccaligner}, for all the methods in the experiments, we calculate the number of reported clone pairs that satisfied the setting of volume difference $\lambda \le$ 0.7. So we can directly compare with their experimental results for fairness. To validate the \emph{FP} (false positive number), for each projects, we randomly selected 100 samples from our results to validate if they are true clone pairs or not and calculated the false positive rate. Then we used the false positive rate to estimate the \emph{FP}.

In order to compare the detection ability of LVMapper with the state-of-the-art tools, we selected the best two clone detection tools for Type-3 and large-gap clone detection SourcererCC and CCAligner. The results data of SourcererCC and CCAligner were taken from that study straightforwardly. We did not provide the result of NiCad here, because NiCad detected almost none of clones with largely different sizes or variances. For all the experiments using these 8 projects, we considered the clones with minimum length of 10 lines, which is consistent with that of the experiments in paper of CCAligner.

The detecting number of large-variance clones (shorted as \emph{LV}) in 8 projects are shown in Table~\ref{tab3}. The number of large variance clones detected by LVMapper was markedly more than that detected by SourcererCC and CCAligner. In project JDK1.2.2, the large-variance clones reported by LVMapper are 970 while CCAligner only reported 15 and SourcererCC only reported 4. CCAligner performed better than SourcererCC at detecting the clones with largely different sizes, which was in fact the target of CCAligner. For all projects we tested on, the precisions (which are $1- \frac{LV}{FP}$) of LVMapper were all over 85\%. Among the reported pairs of LVMapper, we found that many clone pairs has scattered modifications and insertions which were missed by CCAligner. 

\begin{table}[htbp]
\renewcommand\arraystretch{1.2}
\caption{Large-variance Clone Evaluation Results for 8 projects}
\begin{center}
\resizebox{0.9\columnwidth}{!}{%
\scriptsize
\begin{tabular}{ccccccc}
\toprule
\multirow{2}*{Project} & \multicolumn{2}{c}{LVMapper} & \multicolumn{2}{c}{CCAligner} & \multicolumn{2}{c}{SourcererCC}\\
\cline{2-7}
& LV & FP & LV & FP & LV & FP \\
\midrule
JDK 1.2.2 	       & 970 &  87 & 15 & 1 & 4 & 0 \\ 
Ant 1.10.1        & 437 & 56 & 87 & 10 & 13 & 0 \\ 
Maven 3.5.0    & 382 & 34 & 217 & 30 & 38 & 1 \\ 
Opennlp 1.8.1 & 2598 & 78 & 221 & 7 & 5 & 0 \\ 
\midrule
Cook 2.34 	   & 760 & 68 & 63 & 2 & 14 & 0 \\ 
Redis 4.0.0    	& 173 & 26 & 22 & 2 & 7 & 0 \\ 
PostgreSQL 6.0  & 1018 & 102 & 219 & 13 & 38 & 0 \\ 
Linux 1.0 		& 482 & 53 & 27 & 1 & 12 & 1 \\ 

\bottomrule
\end{tabular}
}%
\label{tab3}
\end{center}
\end{table}

We summarized the number of different types clones and the proportion of \emph{LV} clones detected by LVMapper in Table~\ref{tab4}. We classified the clone pairs reported by LVMapper to Type-1\&Type-2, Type-3 and \emph{LV} clones. As we can see from Table~\ref{tab4}, the majority of reported pairs belong to the clones of Type-3. The proportion of the large-variance clones LVMapper reported ranges from 18\% to 62\% in these projects. There is a high proportion of large-variance clones in open source projects and they should not be overlooked.

\begin{table}[htbp]
\renewcommand\arraystretch{1.4}
\caption{Proportion of Large-variance Clones Detected by LVMapper}
\begin{center}
\resizebox{0.95\columnwidth}{!}{%
\scriptsize
\begin{tabular}{cccccc}

\toprule
Project & Type-1\&2 & Type-3 & All  & LV  & LV/ALL \\
\midrule
JDK 1.2.2 & 1102 & 4223 & 5325  & 970 &  18.2\%	\\

Ant 1.10.1 & 114 & 1420 & 1534  & 437 & 28.5\% \\

Maven 3.5.0 & 467 & 1285 & 1752  & 382 & 21.8\%	\\

Opennlp 1.8.1 & 180 & 4006 & 4186  & 2598 & 62.1\%\\

\midrule

Cook 2.34           & 134 & 1847 & 1978  & 760 & 38.4\%  \\ 

Redis 4.0.0         & 41 & 507 & 536  & 173 & 32.3\% \\ 

PostgreSQL 6.0 & 123 & 2141 & 2238  & 1018 & 45.5\% \\

Linux 1.0           & 119 & 1379 & 1493  & 482 & 32.3\%  \\
\bottomrule
\end{tabular}
}%
\label{tab4}
\end{center}

\end{table}

\subsubsection{Large-variance Clone Injection Evaluation}

Considering that the real datasets in previous experiments may be unbalanced in data distribution, we further designed the experiment using self-synthetic data to simulate clones with various proportions of insertions and to measure the impacts on the clone detection tools. Given the source code fragment, we tried to insert scattered lines in different quantities and tested the large-variance detection ability for the clone of the file after insertion with the original one.    

\begin{figure}[htbp]
\centerline{\includegraphics[width=0.5\textwidth]{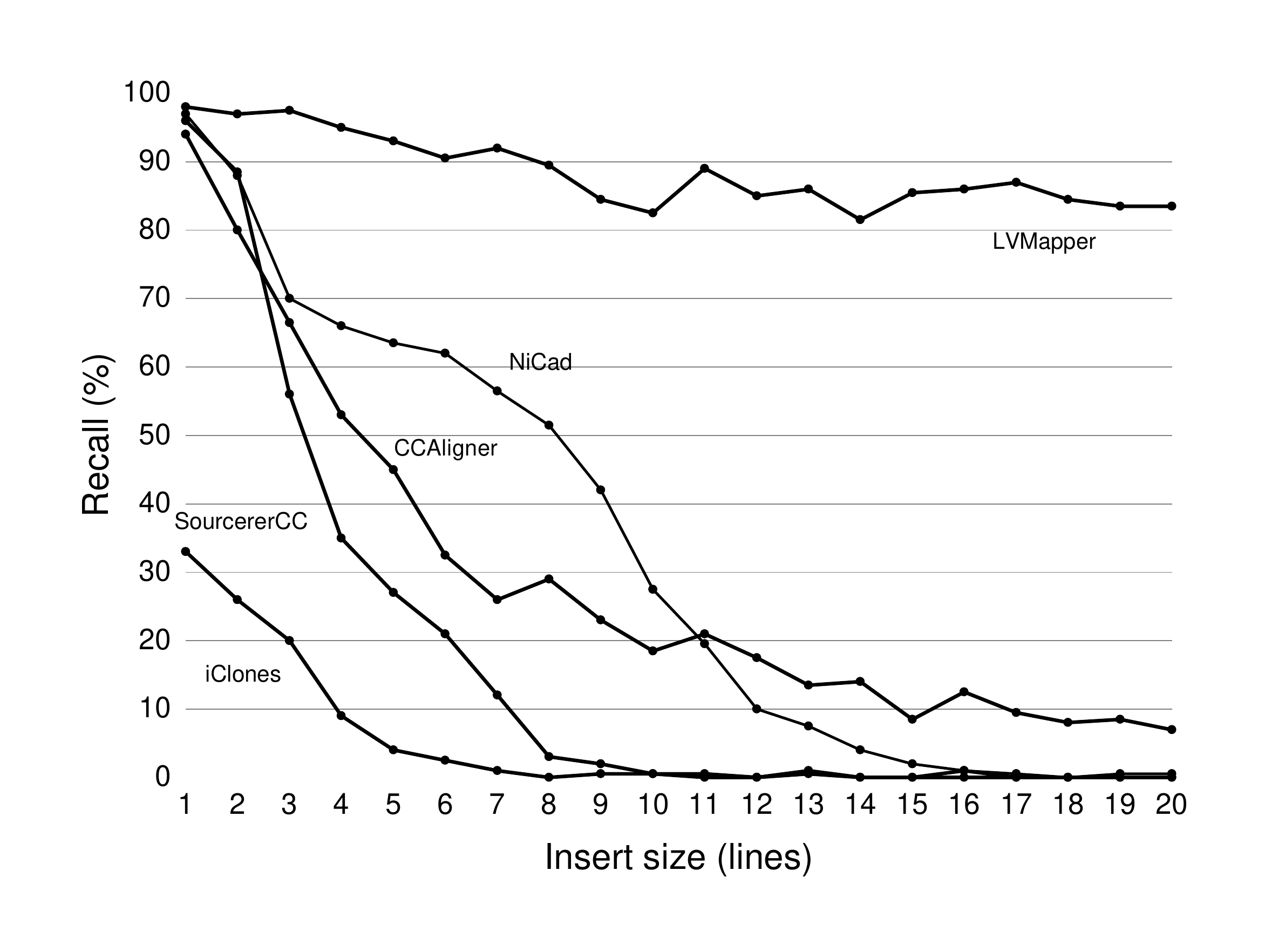}}
\caption{Recall for different insert sizes.}
\label{fig3}
\end{figure}

To evaluate the detection ability of detectors for large-variance clones, we used 200 original code fragments from open source project \emph{jdk 1.8.0} as target code fragments. About one third of them have 15 to 20 lines, one third have 20 to 25 lines, and the remaining part of them have 25 to 30 lines. The number of inserted lines ranges from 1 to 20 lines. For each number of inserting lines, we generated 200 synthetic clones and tested if the tools can detect the clone pairs. We evaluated the state-of-the-art tools CCAligner, SourcererCC, NiCad, iClones with their default configuration. Fig.~\ref{fig3} shows the recall of tools detecting clones for different numbers of inserting lines.

From Fig.~\ref{fig3}, we see that when the number of inserting line is 1, all of the detection tools except iClones had the recall of over 95\%. The recall fell down with the number of inserting lines increasing. When the inserting line was more than 3, SourcererCC, CCAligner and NiCad descended faster while the performance of LVMapper descended slowly and it maintained the highest recall. For number of inserting lines were smaller than 11 NiCad performed better than CCAligner, while CCAligner showed its advantage when the number of inserting lines continued to increases. At the same time, iClones could hardly detect clones when number of inserting lines was more than 7. After insertion of 16 or more lines, the recall of all the other detection tools was lower than 30\% while the recall of LVMapper still remained at above 80\%. In this experiment, NiCad showed good performance when the clone pairs have small variance. CCAligner had the advantage in detecting part of the clones that has relatively concentrated gaps. LVMapper performed the best in both small insertions cases and large-variance cases.

\subsection{General Clone Detection}\label{Bigclonebench}

Apart from the evaluation of detection ability for large-variance clones above, we also compared the detection performance of LVMapper for general clones (i.e. from Type1 to Type3) with other clone detectors. We used BigCloneEval \cite{svajlenko2016bigcloneeval} to test the recall of tools on BigCloneBench \cite{svajlenko2014towards}. The configuration of NiCad was minimum length 6 lines, similarity threshold 70\%, blind renaming and literal abstraction. We used the default configuration of LVMapper, which has seed length of 3-line, and the threshold is described previously. The configuration of SourcererCC was minimum one token and similarity threshold 70\%. We set CCAligner with minimum clone size of 6 lines, window size q = 6, edit distance e = 1, and similarity threshold of 60\%. The result of Deckard \cite{jiang2007deckard}, iClones \cite{gode2009incremental} and CCFinderX \cite{kamiya2008official} were from \cite{sajnani2016sourcerercc}. The number of Weakly Type-3\&4 of Deckard was estimated by the recall rate in \cite{sajnani2016sourcerercc}. As iClones and CCFinderX did not perform well in detecting Moderately Type-3 clones, we did not run them for the performance of Weakly Type-3\&4 (denoted as ``--'' in Table~\ref{tab5}). We evaluated the clone pairs in BigCloneEval with the setting of considering minimum clone size pretty-printed 6 source lines and minimum clone size 50 tokens. For the measurement of precision, as a common practice in the researches \cite{sajnani2016sourcerercc, wang2018ccaligner, saini2018oreo}, we randomly picked 400 pairs from the reported clones of each tool and manually validated the true clone pairs. 

The results for general clone detection performance in BigCloneBench listed in Table~\ref{tab5} has two parts: the last line is the precision and the rest are the recall. The recall of LVMapper for Type-1, Type-2 and Very Strongly Type-3 were 100\% or nearly 100\%. For Strongly Type-3, NiCad performed the best with recall of 95\% and LVMapper was the second with recall of 81\%. With the decreasing in the similarity of the clone pairs, more variance exists within clone pairs. LVMapper had the best performance in Moderately Type-3 with 10\% higher than the second best. The number of Weakly Type-3\&4 clones LVMapper detected was more than double by that of CCAligner. Although Deckard detected the most pairs in Weakly Type-3\&4, it had poor recall for other types of clones and the precision is only 34.8\%. LVMapper kept its precision at about 88\% while showing excellent detecting capability. Compared with the state-of-the-art clone detectors, LVMapper has good recall and precision for all types of clones.

\begin{table*}[htbp]
\renewcommand\arraystretch{1.3}
\caption{Recall Per Clone Type and Precision Measured for BigCloneBench}
\begin{center}
\resizebox{1.5\columnwidth}{!}{%
\scriptsize
\begin{tabular}{ccccccccc}
\toprule
Type   & LVMapper  & CCAligner   & NiCad & SourcererCC & Deckard & iClones & CCFinderX \\
\midrule
Type-1                        & 100   	  & 100     & 100   	& 100   & 60	  & 100	  & 100	\\

Type-2                        & 99      & 99          &100    	& 98	   & 58      & 82 	          & 93		\\

Very Strongly Type-3 & 98       & 97         & 100         & 93	   & 62	  & 82 	  & 62		\\

Strongly Type-3         & 81      & 70           & 95         & 61	  & 31	  & 24 	  & 15		\\

Moderately Type-3     & 20     & 10           & 1	      & 5       & 12 	  & 0 	          & 1		\\
	
\# of Weakly Type-3\&4 & 30250 & 12540  & 12  & 1892 & 77293       & -- 	         & --		\\

\midrule

Precision                 & 88      & 78.8       & 94.5     & 98.8  	  & 34.8    & 91 	        & 72 	         \\ 

\bottomrule
\end{tabular}
}%
\label{tab5}
\end{center}
\end{table*}

\subsection{Comparison with semantic method}\label{Comparison}

In the experiments above, the detection ability of LVMapper and other non-semantic tools were tested thoroughly. In order to compare the clone detection ability of LVMapper with semantic method and analyze the difference between them, here we compared LVMapper with the latest machine learning based tool Oreo \cite{saini2018oreo}. Because Oreo only supported clone detection with Java code, we used the same Java projects mentioned in Section \ref{LvDetect}. Oreo was executed with the default configuration.

Table~\ref{tab7} shows the \emph{LV} (number of large-variance clones) detected by LVMapper and Oreo for the Java projects and \emph{O/L} is the ratio of \emph{LV} that Oreo could detect among the \emph{LV} detected by LVMapper. The results shows that the large-variance clones detected by Oreo are only a small part of that detected by LVMapper. The main reason is that the semantic clone detector focuses on detecting clones that are almost the same or very similar in semantic.  As the semantics of code with more modifications may be changed, the large-variance clones are difficult to be detected by semantic methods.

\begin{table}[htbp]
\renewcommand\arraystretch{1.2}
\caption{Large-variance Clones Number for 4 java projects}
\begin{center}
\resizebox{0.85\columnwidth}{!}{%
\scriptsize
\begin{tabular}{cccccc}
\toprule
\multirow{2}*{Project} & \multicolumn{2}{c}{LVMapper} & \multicolumn{2}{c}{Oreo} & \multirow{2}*{O/L}\\
\cline{2-5}
& LV & All & LV & All &\\
\midrule
JDK 1.2.2 	        & 970 & 5325 & 229  & 3553 & 23.6\%\\ 
Ant 1.10.1         & 437 &1534 & 143  & 1941 & 32.7\%\\ 
Maven 3.5.0      & 382 & 1752 & 121 & 1441& 31.7\%\\ 
Opennlp 1.8.1   & 2598 & 4186 & 151 & 1775 & 5.8\%\\ 

\bottomrule
\end{tabular}
}%
\label{tab7}
\end{center}
\end{table}

\subsection{Scalability}

To test the scalability of LVMapper, we selected 1M LOC, 10M LOC, 20M LOC and 30M LOC from the inter-project Java repository IJaDataset-1.0 \cite{ambient11} as the target files to detect clones. We used a standard desktop with a 3.5GHz quad-core i7-4770k CPU and 24GB of memory. As CCAligner and SourcererCC had relative good scalability in a recent study \cite{wang2018ccaligner}, we compared the execution time of LVMapper with CCAligner and SourcererCC. The minimum lines was set as 6 for LVMapper and CCAligner, and the minimum tokens was set as 50 for SourcererCC.

The execution time across different scales of datasets are shown in Table~\ref{tab6}. SourcererCC had less execution time for 1M LOC to 30M LOC and it ran faster than LVMapper on 30M LOC. CCAligner scaled to 10M LOC and it failed for the 20M LOC and 30M LOC inputs with an out of memory error (denoted as ``--'' in Table~\ref{tab6}). LVMapper was the fastest for 1M LOC, 10M LOC and 20M LOC. For 30M LOC, LVMapper was slightly slower than SourcererCC but they were of the same order of magnitude. Although SourcererCC has good scalability but the large-variance detection ability of SourcererCC is limited. 
 
\begin{table}[htbp]
\renewcommand\arraystretch{1.2}
\caption{Execution time for different LOC}
\begin{center}
\resizebox{0.85\columnwidth}{!}{%
\scriptsize
\begin{tabular}{cccccccc}
\toprule
LOC  	 	&1M 		&10M  		& 20M    		& 30M  			\\
\midrule
LVMapper	          & 34s  & 22m 40s            &  1h 22m 2s    	&  3h 11m 18s		\\

CCAligner 	& 41s  & 1h 1m 40s   	&   --       	 	& --				\\
	
SoucererCC     & 4m 48s  & 31m 12s 	&  1h 32m 21s   &  2h 18m 50s		\\

\bottomrule
\end{tabular}
}%
\label{tab6}
\end{center}
\end{table}

\section{Related Work}\label{Related Work}
There are many code clone detection tools proposed in the literature. More descriptions of these tools and methods can be found in \cite{burd2002evaluating, koschke2007survey, roy2007survey, bellon2007comparison, roy2009comparison, rysselberghe2004evaluating, arcelli2013software, rattan2013software, sheneamer2016survey}. At present, the code clone detection of Type-3 is still a difficult task, especially for large-variance code clones. According to the types of clone similarity, the clone detection methods can be divided into two categories. One is the non-semantic method, and the other is the semantic method. Our method belongs to the former.

\subsection{Non-semantic Methods}
These methods or tools determine whether the code pairs are clones or not base on the similarity of code words and code sentences. These clone detection methods mainly include the text based, the token based, the tree and graph based and the metrics based methods. Among these methods, some researchers classified the latter two as the semantic method.

For the text based tools \cite{johnson1994substring, ducasse1999language, roy2008nicad}, two code blocks are compared in the form of text or strings. Johnson \cite{johnson1994substring} proposed a fingerprinting technique to identify similar source code and to speed up processing speed. Ducasse \cite{ducasse1999language} developed a line based comparison detection tool. NiCad \cite{roy2008nicad} is based on a two phases process, viz., identification of potential clones and code comparison using longest common subsequences. Compared with LVMapper, it adopts similar locating and verifying strategy. NiCad can detect Type-3 clones, but did not perform well in the test of detection ability for large-variance clones as shown in Fig.~\ref{fig3}.

For the token based tools \cite{kamiya2002ccfinder, gode2009incremental, sajnani2016sourcerercc}, tokens are firstly extracted from the source code by lexical analysis, and it is better than simple keyword matching since it tolerates different identifiers. CCFinder \cite{kamiya2002ccfinder} is a popular tool based on token, but it does not support Type-3 clone detection. In their work, they used suffix tree to find identical subsequences and increase the threshold to filter small clones. Essentially, these operations are equivalent to the indexing and filtering technology in LVMapper. iClones \cite{gode2009incremental} and SourcererCC \cite{sajnani2016sourcerercc} are also influential representatives of such tools. G{\"{o}}de and Koschke \cite{gode2009incremental} developed the incremental tool iClones by merging neighboring Type-1/Type-2 clones to big clones or Type-3 clones. However, iClones can only detect Type-3 clones with small variance. Sajnani \cite{sajnani2016sourcerercc} developed a fast clone detection tool SourcererCC which uses tokens composition to verify clones, but it is constrained to the identification ability of token granularity. CCAligner \cite{wang2018ccaligner} has good performance in detecting clones with relatively concentrated modifications but it misses scenarios where modifications are scattered.

For the tree and graph based tools \cite{yang1991identifying, jiang2007deckard, krinke2001identifying, komondoor2001using, wang2017ccsharp}, abstract syntax tree (AST) and program dependency graph (PDG) are frequently used as the representation of source code. Yang \cite{yang1991identifying} and Deckard \cite{jiang2007deckard} proposed AST approaches for finding the syntactic differences between two programs. Duplix \cite{krinke2001identifying} and PDG-DUP \cite{komondoor2001using} are PDG-based tools which use program slicing to find isomorphic subgraphs. These tree and graph based tools suffer from large execution times and poor scalabilities. To this end, CCSharp \cite{wang2017ccsharp} improves the time performance and accuracy of the PDG-based method, but it still cannot achieve good performance in large scale dataset. Besides, these tools will fail to detect large-variance clones since structure of tree and graph may be changed during the extension and modification of the code.

For the metrics based tools \cite{mayrand1996experiment, balazinska1999measuring, patenaude1999extending}, some metrics and characteristic features for tree and graph of source code can be used for code clone detection. Both Mayrand \cite{mayrand1996experiment} and Balazinska \cite{balazinska1999measuring} extracted metrics from an AST representation of source code and used the metrics for clone identification. Patenaude \cite{patenaude1999extending} used the metrics of source code that can be divided into five categories, viz., classes, coupling, methods, hierarchical structure and clones. These methods extract some features from tree or graph or source code to verify the semantic similarity of two code blocks. They have similar limitations to that of the tree and graph based tools for large-variance clones.

\subsection{Semantic Methods}
Apart from the tree and graph based and the metrics based tools mentioned above, these kind of clone detection tools include the semantic space mapping based tools \cite{marcus2001identification}, the software behavior based tools \cite{choi2009static, jiang2009automatic, kim2011mecc} and so on. Substantially, the tools based on semantics adopt semantic abstraction or modeling for source code rather than abstraction of lexical and syntactic similarity. Due to overlap of semantic clones and large-variance clones, however, the methods based on semantics can also find a small part of large-variance clones shown as Table~\ref{tab7}.

Machine learning is always an effective way to deal with complex problems, including code clone detection especially for semantic clone. White \cite{white2016deep} presented an unsupervised deep learning approach to detect clones, which can automatically learn discriminating features of source code. Wei \cite{wei2017supervised} proposed a method to detect clones by learning representations and Hamming distance of code fragments. Zhao \cite{zhao2018deepsim} encoded code control flow and data flow into a semantic matrix for detecting semantic clones. Recently, Saini \cite{saini2018oreo} put forwarded a machine learning based method called Oreo, which can find the code clones in the overlap between syntactic and semantic zone. Oreo used the clones that are almost identical or very similar to train the deep learning model and the large-variance clone detection ability of Oreo is limited. Machine learning methods always face the issues of efficiency and dependency on the initial training data. The experiment in Section \ref{Comparison} shows that the machine learning method could not detect the large-variance clones well, which means the modifications between clones may change the semantics of code.

\section{Limitation}\label{Limitation}
Our tool is aimed at detecting the large-variance clones with \emph{homologous modification}. As shown in Fig.~\ref{fig3} for the large-variance clone injection evaluation, the large-variance clone detection ability of LVMapper is significantly better than others. Note that this is the experiment of large variance clone detection just for syntactic similarity and our method did not detect semantic clones. In the BigCloneBench experiment shown as Table~\ref{tab5}, our tools have made better progress in Moderately Type-3 and Weakly Type-3\&4, but it has not yet reached a satisfactory level. The reason is probably that Moderately Type-3 and Weakly Type-3\&4 clones are more of semantic clones that are generated by \emph{heterologous development}.

The scalability of LVMapper on larger scale dataset needs to be tested. LVMapper on 30M LOC shows satisfactory performance (3h) in the evaluation. It is meaningful to test it on over 100M LOC for validating its scalability further.

\section{Conclusion \& Future Work}\label{Conclusion}
The large-variance code clones are generated by homologous modification and can be used in software development and other applications. Our experiments found that these clones were widespread in Type-3 clones, even in some datasets up to half or more. And the large-variance code clone changes the past clone detection methods that focus on finding almost identical or very similar code pairs. Therefore, The research on large-variance code clone is important and meaningful. In this paper, we proposed a novel concrete definition and a detector LVMapper borrowing from the idea of sequencing alignment in bioinformatics for large-variance code clones. Effective and innovative technologies such as dynamic threshold, avoiding time-consuming dynamic programming and seeds index are designed in our method. A series of testing on real cases of software projects, self-synthetic programs and the state-of-the-art benchmarks showed the large-variance clone detection performances of LVMapper are much better than the other state-of-the-art tools, and it has good recall and precision for general Type-1 to Type-3 clones. Furthermore, we will make LVMapper more scalable for the clone detection on larger scale datasets. And it will be important work to do research on software engineering applications such as code recommendation and completion, refactoring and bug propagation for large-variance clones in the future.

\bibliographystyle{IEEEtran}
\bibliography{sample-base.bib}

\end{document}